\documentclass[seceq]{ptptex}

\usepackage{graphicx}
\usepackage{color}

\title{
Synchronisation Induced by Repulsive Interactions in a System of van der Pol Oscillators%
}

\author{
Teresa \textsc{Vaz Martins}$^{1,2}$ and Ra\'ul \textsc{Toral}$^{1}$%
}

\inst{
(1)IFISC, Instituto de F{\'\i}sica Interdisciplinar y Sistemas Complejos, CSIC-UIB,  Campus UIB, E-07122 Palma de Mallorca, Spain \\(2)Centro de F{\'\i}sica do Porto, DFA, FCUP, 4169-007 Porto, Portugal
}

\abst{
We consider a system of identical van der Pol oscillators, globally coupled through their velocities, and study how the presence of competitive interactions affects its synchronisation properties. We will address the question from two points of view. Firstly, we will investigate the role of competitive interactions on the synchronisation among identical oscillators. Then, we will show that the presence of an intermediate fraction of repulsive links results in the appearance of macroscopic oscillations at that signal's rhythm, in regions where the individual oscillator is unable to synchronise with a weak external signal. 

}

\begin{document}

\maketitle

\section{Introduction}
\label{intro}

 Synchronisation,\cite{pik} or the ability of coupled oscillators to adjust their rhythms, is a property that arises in many systems, from pacemaker cells in the heart firing simultaneously as a result of their interaction,\cite{pace} to the fetal heart rate adjusting its pace to maternal breathing, as an example of  forced synchronisation.\cite{fet}

Typically, oscillators with different frequencies are able to synchronise due to a strong enough positive coupling among units. However, interactions in Nature are often repulsive and, surprisingly,  it was found that under some particular circumstances repulsive interactions can actually enhance synchronisation: thus, the presence of negative links can prevent the instability of the fully synchronised state when it compensates an excessive number of positive links,\cite{pnas} or its sparse presence can enhance synchronisation in small-world networks.\cite{Levya} \ Most interestingly - since it is not always desirable to achieve a state of full synchronisation - the presence of repulsive links can give rise to new forms of synchronisation,\cite{tsim} that sometimes can be described as glassy or glassy-like.\cite{daido1, daido2, daido3, glassy2} \ Additionally, the beam-forming abilities of a system of repulsively coupled Stuart-Landau oscillators were considered in.\cite{beam} 

So far, studies have mostly focused on non-identical phase oscillators, and several coupling schemes have been chosen, such as local \cite{preprint} or long-range,\cite{Levya} and purely repulsive \cite{tsim} or assuming a competition between repulsive and attractive.\cite{zan} \ Like in,\cite{zan} we want to isolate the effect of different proportions of repulsive interactions by considering identical oscillators. However, rather than establishing how full synchronisation becomes unstable as the fraction $p$ of repulsive links increases,\cite{zan} our focus will be on the characterisation of the different configurations that emerge as $p$ grows, and its implications for signal transmission when the system is subjected to an external forcing.   Also, unlike \cite{zan} we will not consider phase oscillators, but instead van der Pol oscillators,\cite{vdp} which implies phase, amplitude and frequency synchronisation are taken into consideration. 

The establishment of the role of the coupling structure on synchronisation, independently of the detailed specification of the nodes dynamics, can rely on the study of the Laplacian matrix.\cite{msf, pat} \ We will identify and characterise a transition region from synchronisation to desynchronisation by analysing the eigenmodes of the Laplacian matrix corresponding to different proportions of repulsive links, adapting a formalism developed in.\cite{fi4, Perc}  

The second part of the paper will be devoted to explore the role of competitive interactions in the synchronisation of the system with an external periodic signal. We will choose a signal whose frequency lies outside the region of entrainment for an uncoupled oscillator, as well as for an all attractively coupled system. This problem is closely related to a second theoretical framework, that of resonance studies, that emphasise the importance of an intermediate disorder on the response to an weak signal, where disorder can be  noise,\cite{G, HM} diversity,\cite{TMTG06, TTVL, linosc} or competitive interactions  \cite{fi4, dac}. In the latter cases \cite{fi4, dac} it was found that an intermediate fraction of repulsive links was able to amplify the response to an external signal, in bistable systems where the external signal was be the only source of movement.  In the present case, an optimal response should correspond to an adjustment between the intrinsic frequency of the oscillators and that of the external signal; as we will see, that optimal response is achieved at an intermediate proportion of repulsive links in the case  of strong fast signals, whereas weak slow signals are best responded when all the links are negative.

The outline of this paper is as follows: in section \ref{model} we will introduce the model; we show that an increase of the proportion of repulsive links leads to loss of synchronisation in section \ref{des}; and in section \ref{response} we show how the presence of repulsive links accounts for an enhanced response to external signals; in section \ref{ext} we will briefly mention some extensions; and conclusions are drawn in section \ref{conclusions}.

\section{Model}
\label{model}

We consider an ensemble of van der Pol oscillators \cite{vdp} $\{x_i(t),i=1,\dots,N\}$, globally coupled through their velocities $\dot x_i$, and subjected to an external periodic forcing of amplitude $A$ and frequency $\Omega$. The dynamics is described by:

\begin{equation}\label{vdp}
\ddot{ x_i } =  - x_i + \mu(1 - x_i^2)\dot{ x_i } + \frac C N \sum_{j=1}^N J_{ij}\left( \dot{x_j}- \dot{x_i}\right) + A\sin (\Omega t),
\end{equation}
where the nonlinearity parameter $\mu$ is a positive constant and $C$ is the coupling strength.

The coupling between the oscillators $i$ and $j$ is given by the interaction term $J_{ij}$ and can be attractive or repulsive, according to a given probability $p$.

\begin{equation}
J_{ij}=J_{ji}= \begin{cases}
 -1, & \text{with probability $p$},\\
 1, & \text{with probability $1-p$}.
\end{cases}
\label{Eq:Jij}
\end{equation}

The single van der Pol oscillator is a paradigmatic example of a non-linear oscillator. It possesses a stable limit cycle as a result of its nonlinear damping term $\mu(1 - x_i^2)$: for small oscillations, $|x_i|<1$, the system experiences negative damping and the oscillations grow, while for $|x_i|>1$, the positive damping causes the oscillations to decrease.  Therefore, independently of the initial conditions or small perturbations, its amplitude of oscillations reaches a constant (equal to 2), while its detailed shape and period $T$ depend on $\mu$, approaching $T \approx(3-2 \ln 2)\mu$ for large $\mu$.   In this case of large $\mu\gg 1$, the oscillations are called relaxational and are characterised by the presence of discontinuous jumps intercalated by periods of slow motion.

\section{Desynchronisation among the unforced oscillators, $A=0$}
\label{des}

\begin{figure}
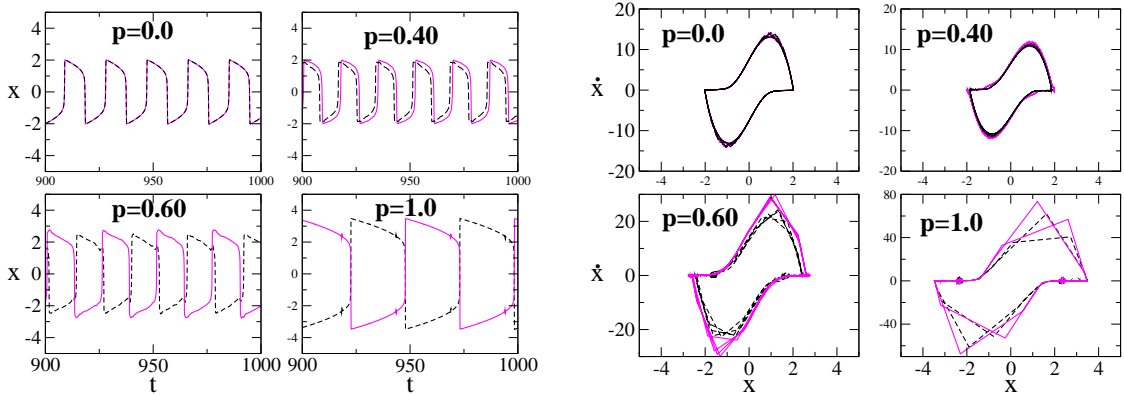

\vspace{0.5cm}
\includegraphics[width=7cm,angle=0]{utraj.eps}
\hspace{0.5cm}
\includegraphics[width=7cm,angle=0]{fase.eps}
\caption{Trajectories (left panels) and phase portraits (right panels) of two individual oscillators, for various probabilities $p$ of repulsive links.  Regardless of initial conditions, the van der Pol oscillator reaches steady state oscillations with a constant amplitude. In the right panel we represent the corresponding limit cycles in the phase plane of $(x, \dot x)$: as $p$ grows the distinction between slow and fast motion becomes clearer which is manifested in the more abrupt angles in the limit cycles for $p=0.6$ and $p=1.0$. N=100, C=20, $\mu=10$, A=0.  
}
\label{tra}
\end{figure}

Fig.~\ref{tra}  shows the trajectory (left panels) and respective limit cycles (right panels)  of two typical individual oscillators for some probabilities $p$ of repulsive links. In all cases, the essential characteristics of the van der Pol oscillator, such as a steady amplitude and the existence of two time scales, are preserved by this type of coupling, as it is reflected by the fact that the stable limit cycles  (Fig.~\ref{tra}, right panels)  maintain its basic shape.   Having identical natural frequencies, when the coupling constant $C$ is strong enough the position of the oscillators becomes synchronised when all the interactions are attractive. (Fig.~\ref{tra}, for $p=0$).   As the proportion of repulsive links grows, both the amplitude and the phase of the oscillators start to desynchronise (Fig.~\ref{tra}, $p=0.4$). Finally, a further increase in the proportion of repulsive links (Fig.~\ref{tra}, $p=0.60$ and $p=1.0$), drives the system to a configuration where the global variable $X(t)=\frac{1}{N}\sum_i x_i(t)$ is zero, with several groups oscillating in anti-phase, with  a decreased frequency and an increased amplitude of oscillations: if we consider two groups with the same size that are oscillating in anti-phase, it can be shown that their amplitude becomes rescaled as a result of their interaction by a factor of $\sqrt{1 + \frac{C}{\mu}}$.

We can describe the last configuration characterised by a zero value of the average position $X(t)$ as a disordered situation at the macroscopic level. To quantify this \emph{disordering} role of repulsive links, we define, following,\cite{linosc} the complex variable $z_i=x_i + i \dot x_i$, the average $\bar z=\frac{1}{N}\sum_{i=1}^N {z_i}$ and the  variance of $z_i$  normalised by the average value of the modulus squared, $\sigma^2[z_i]$:

\begin{equation}
\sigma^2[z_i]=\left\langle\frac{N^{-1}\sum_{i=1}^N |z_i-\bar z|^2}{N^{-1}\sum_{i=1}^N |z_i|^2}\right\rangle, 
\end{equation}
here and henceforth $\langle\cdots\rangle$ denotes a time average.

The normalised variance can take values between  $\sigma^2[z_i]=1$ for maximum disorder, and   $\sigma^2[z_i]=0$ when all oscillators are synchronised amongst themselves. From this, we choose \cite{linosc} a measure of order that reduces to the Kuramoto order parameter \cite{kura} when all units oscillate with the same amplitude:

\begin{equation}
\label{eq:rho}
\rho=\sqrt{1-\sigma^2[z_i]}.
\end{equation}

\begin{figure}
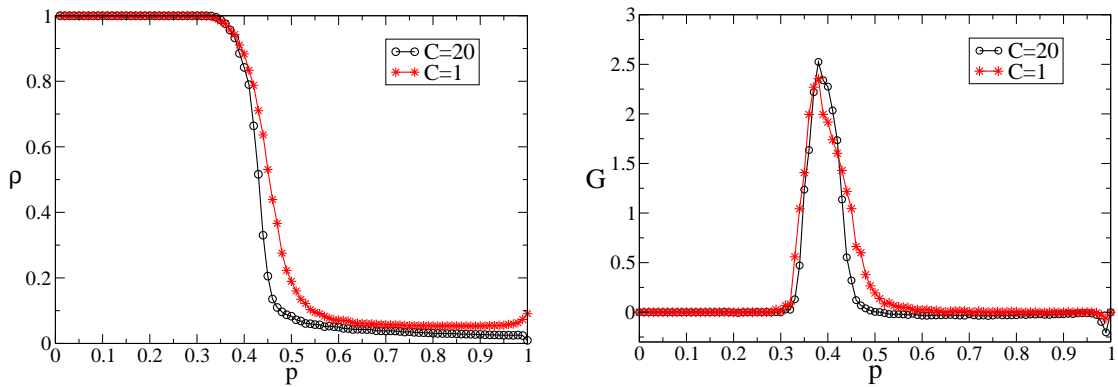

\vspace{0.5cm}
\includegraphics[width=7cm,angle=0]{dis.eps}
\hspace{0.5cm}
\includegraphics[width=7cm,angle=0]{g.eps}
\caption{The measures of disorder $\rho$ (left panel) and $G$ (right panel). Averages over 100 runs, and N, C, $\mu$ and A as in Fig. \ref{tra}}.
\label{dis}
\end{figure}

As dispersion increases, $\rho$  decreases from $\rho=1$ to $\rho=0$. As we show in Fig \ref{dis}, there is a clear synchronisation-desynchronisation transition for an intermediate fraction of repulsive links, that does not depend much on the coupling strength $C$.

To characterise desynchronisation further, it is useful to look into the behaviour of the field that an oscillator feels as a result of the interaction with other units. The average number of {\sl effective links} $F=\frac{1}{N^2}\sum_{ji} J_{ij}=\frac{1}{N^2}(2p-1)(N-1)$,  in a given run can in general be different from the particular number an oscillator has, $f_i=\frac{1}{N} \sum_{i=1}^N J_{ij}$, given that the probability of repulsive links $p$ follows a binomial distribution with the corresponding variance. 
 We want to know if there is a correlation between the fraction of repulsive links an oscillator has and its synchronisation with the overall majority. That is described by the following quantity $G$:

\begin{equation}
G = \left\langle \frac{1}{N} \sum_{ji}\dot{x_j} \dot{x_i} \left[f_i - F \right]\right\rangle = \left\langle\dot{X} \sum_{i}\dot{x_i} \left[f_i - F \right]\right\rangle, 
\label{g}
\end{equation}
being $\dot{X}= \frac{1}{N} \sum_{j}\dot{x_j}$ the mean velocity.
We observe that the order-disorder transition region $p\sim[0.4, 0.45]$ we identify in the left panel of Fig. \ref{dis}, is accompanied by an increase in the influence on an oscillator of its particular coupling configuration, as signalled by the peak in $G$ (Fig. \ref{dis}, right panel). The oscillators with a higher than average number of repulsive links form a loosely synchronised group in a different slow region than the one where the majority concentrates.

The partial  independence of the state of the oscillator on the global configuration opens the possibility of the existence of several different global states, and thus hints at the existence of metastable states.  We can relate this behaviour to the coupling structure, by the spectral analysis of the associated Laplacian matrix $J'_{ij}=J_{ij}-\delta_{ij}\sum_{k=1}^N J_{kj}$,\cite{fi4} where $\delta_{ij}$ is the Kronecker's delta.    We  begin by  rewriting Eq. (\ref{vdp}) as system of two equations that highlight a fast motion for the $x_i$ variable and a slow motion for the $y_i$ variable:

\begin{subequations}
\begin{eqnarray}
\dot{ x_i } & = & \displaystyle   \mu \left[x_i - \frac{1}{3}x_i^3 - y_i + \frac D N \sum_{j=1}^N J'_{ij}x_j\right] \label{li1}\\
\dot{ y_i } & = & \displaystyle   \frac{1}{\mu} \left[x_i - A \sin (\Omega t)\right]. \label{li2}
\end{eqnarray}
\end{subequations}
where $D = \frac{C}{\mu}$. 

We focus on Eq. (\ref{li1}), letting the slow variable $y_i$ be a constant.  The Laplacian appears naturally in the coupling term, and  we can see its positive eigenvalues should help deviations from a given state along the $x$ variable.\cite{fi4, pat} \ We now introduce the   eigenvalues $Q_{\alpha}$ and eigenvectors $e^{\alpha}=(e_1^{\alpha},\dots,e_N^{\alpha})$ of the Laplacian matrix, with the normalisation condition $\sum_i e_i^{\alpha}e_i^{\beta} = \delta _{\alpha \beta}$.

\begin{equation}
\label{eig}
\sum_{j=1}^NJ'_{ij}e_j^{\alpha}=Q_{\alpha}e_i^{\alpha}.
\end{equation}

Let's us assume the state of a unit $i$ is $x_i^o$ at a given time, where $x_i^o$ is drawn from any symmetric random distribution. We perturb the initial states as  $x_i^o+ s_i$, and express $s_i$ in terms of the eigenvalues and eigenvectors of the Laplacian, so that $s_i=\sum_{\alpha=1}^NB_{\alpha}e_i^{\alpha}$. We aim to see how the interaction with other units influences the reaction to perturbations.  

 We will assume for simplification that $\sum_{i=1}^Nx_i^oe_i^{\alpha} \sim \sum_{i=1}^Nx_i^o\sum_{i=1}^Ne_i^{\alpha}=0$, and that the modes are uncoupled. After expanding the equation in terms of the Laplacian eigenvalues and eigenvectors, we then multiply the resulting equation by $e_i^{\alpha}$ and average over all elements, to find the evolution for the amplitude of the $\alpha$-th mode:

\begin{equation}
\label{mode}
\frac{dB_{\alpha}}{dt}=-\frac{1}{3}B_{\alpha}^3 +\text{PR}_{\alpha}\left(\frac{C}{\mu}\frac{Q_{\alpha}}{N}-k\right)B_{\alpha}, 
\end{equation}
where $k=\frac{1}{N} \sum_{i=1}^N (x_i^o)^2 -1$ is a quantity related to the variance of the initial conditions,  and the participation ratio $\text{PR}_\alpha=1/\sum_{i=1}^N [e_i^{\alpha}]^4$ is a classical measure of localisation that estimates the number of oscillators that participate significantly in a state $e^{\alpha}$: for a state localised on a fraction $f$ of elements, $\text{PR}_\alpha$ tends to $f$. According to Eq. (\ref{mode}), unless $Q_{\alpha}>\frac{k N \mu}{C}$, the amplitude of the mode $B_{\alpha}$ tends to zero, and any deviation from the initial state vanishes. Otherwise, mode $\alpha$ is said to be an {\sl open mode}. 

\begin{figure}
\vspace{0.5cm}
\centerline{\includegraphics[width=7cm,angle=0]{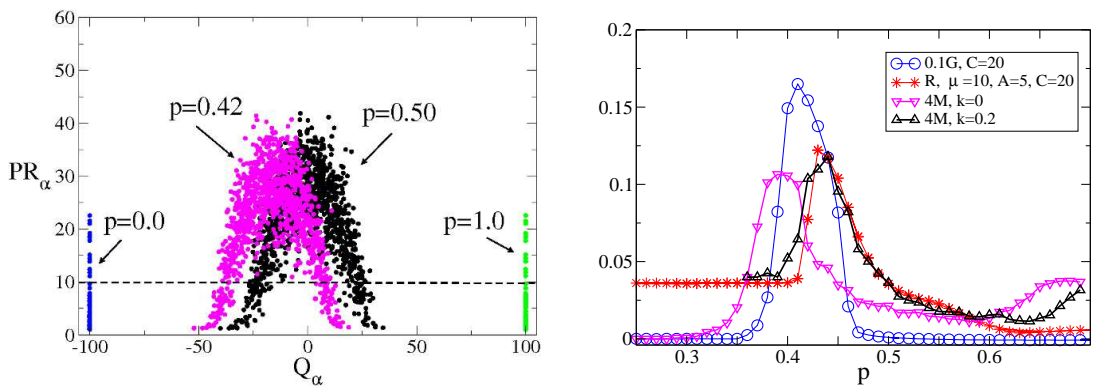}
\hspace{0.5cm}
\includegraphics[width=6.5cm,angle=0]{gmr.eps}}
\caption{Left panel: we plot the participation ratio $PR_\alpha$ for the $Q_\alpha$ eigenvalues. As the probability of repulsive links $p$ grows the eigenvalues become dislocated towards higher values.  The eigenvalues whose participation ratio is below the dashed line are localised. Right panel: the measures of localisation M, the measure of disorder G, and the spectral power amplification R, all have the maximum value at roughly the same p. For better viewing, we multiplied M by 4, and G by 0.1 N=100. In the case of M and k=0.2 we have C=20 and $\mu=10$.
}
\label{comp}
\end{figure}

In a precise way, we will define ``localised'' modes as the ones whose participation ratio is less than $0.1N$, and define a measure $M$ of localisation \cite{fi4} as $M=\frac{N_L^2}{N_O N}$, where $N_L$ is the number of open localised modes, i.e. those satisfying $\text{PR}_{\alpha}<0.1N$, and $N_O$ is the total number of open modes $\alpha$. While for extreme probabilities $p$ of repulsive links the number of possible values for $Q_\alpha$ is very restricted, we find that for intermediate levels of $p$ the distribution of $Q_\alpha$ is broader and the eigenvalues at both tails of the spectrum are localised: the left panel of Fig. \ref{comp} illustrates this fact for some examples of $p$.

 In our regime of strong non-linearity, or $\mu\gg1$, the time an oscillator spends on a fast motion is very close to zero; in such case we should have $k\approx0$ and the condition for open modes can be fulfilled at low enough probabilities of repulsive links.  Of the open modes, those that correspond to positive Laplacian eigenvalues facilitate the growth of perturbations, while those that are negative inhibit it. We wish to identify a probability of repulsive links that makes the system flexible enough to sustain deviations from the initial symmetric configuration, yet stable enough so that perturbations don't spread immediately throughout the entire system, as that is what we observe at the transition region (Fig. \ref{dis} right panel), where some oscillators are more loosely synchronised than others. We can anticipate \cite{fi4} that such situation corresponds to a localisation \cite{and58, and70} of positive Laplacian modes.

As we see in Fig. \ref{comp}, the peak in $G$, that signals the range of $p$ where there is an heightened dependence of the state of an oscillator on its coupling structure (Fig. \ref{dis}), coincides with a localisation of the positive eigenvalues of the Laplacian. Those localised positive open modes are responsible for keeping the loss of synchronisation within controllable limits.

\section{Synchronisation with the external signal, $A\neq 0$}
\label{response}

In this section, we will see how competitive interactions affect the response to an external periodic signal.  Since in general there can be several frequencies present in the output of the global variable $X(t)=\frac{1}{N}\sum_i x_i(t)$, we say the system is synchronised with the external signal when the highest peak in the Fourier spectrum corresponds to that frequency. 

When the natural frequency of oscillations coincides with the external forcing frequency, synchronisation is achieved for vanishing $A$, and as the two frequencies diverge, stronger forcings are needed to entrain the system. We will call a signal \emph{strong} when its amplitude is greater than the amplitude of oscillation of the unforced van der Pol unit, and we will call it \emph{fast} when its frequency is higher than the natural frequency of the individual van der Pol oscillator.

We will distinguish between strong fast and weak slow signals, because the mechanism of synchronisation differs in the two cases, although, in both cases, competitive interactions are required for an enhanced response.

\subsection{Strong fast signals benefit from intermediate $p$}
\label{fast}

\begin{figure}
\vspace{0.5cm}
\centerline{\includegraphics[width=6cm,angle=-90]{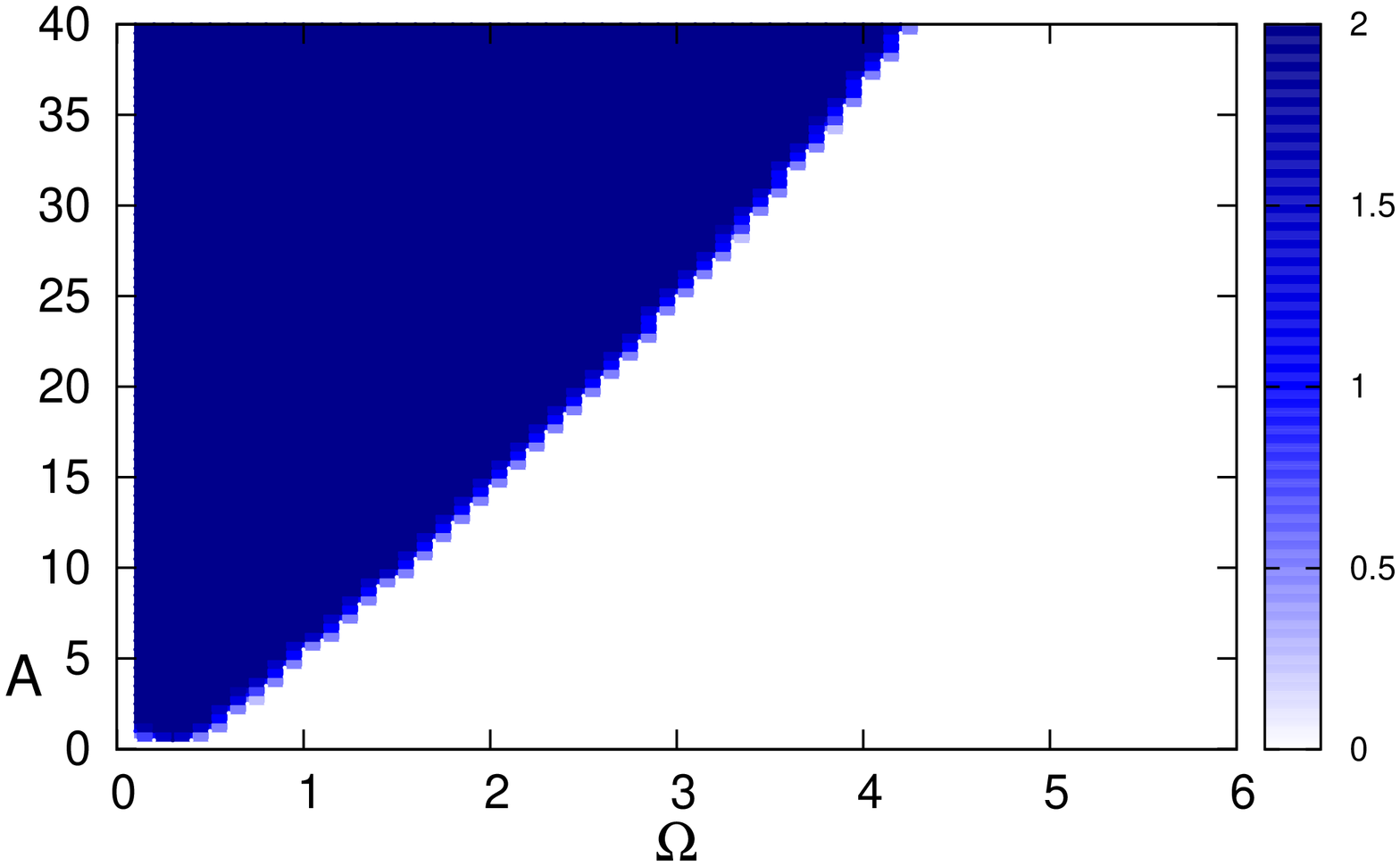}
\includegraphics[width=6cm,angle=-90]{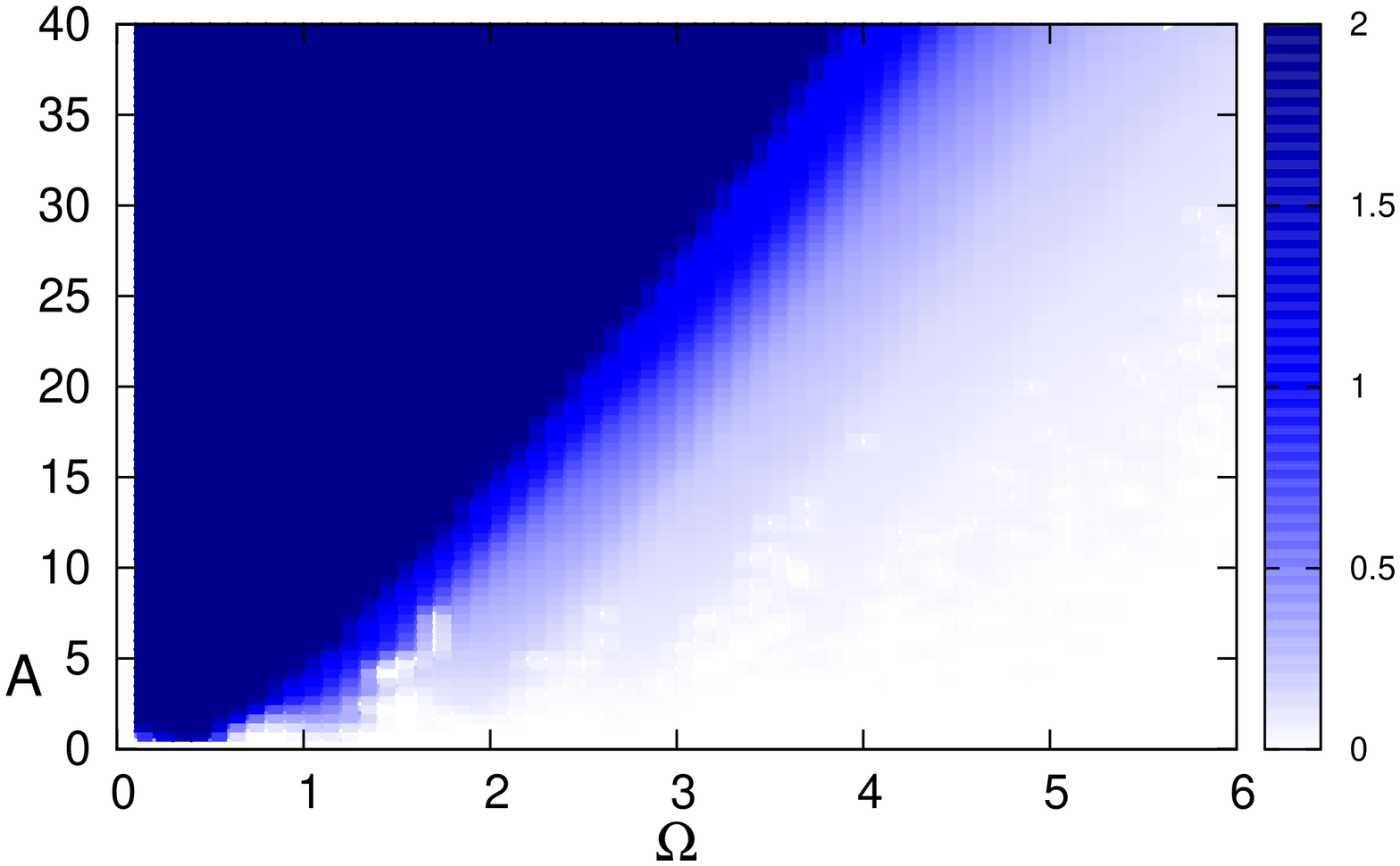}}
\caption{We plot the spectral power amplification $R$ in the synchronisation regions for $p=0$ (left panel) and $p=0.43$ (right panel). For better viewing, we use a color codes that saturates for $R\ge2$. N, C and $\mu$ as in Fig. \ref{tra}. 
}
\label{arn}
\end{figure}

In Fig. \ref{arn} we plot the synchronisation regions and their relative strength, as measured by the spectral power amplification factor~\cite{ampli} $R$, given by:

\begin{equation}
R=4A^{-2}\left|\langle{\rm e}^{-i \Omega t}X(t)\rangle\right|^2
\label{R}
\end{equation}
$R$ is roughly proportional to the square of the normalised amplitude of the oscillations of $X(t)$ at the frequency $\Omega$, being $R<1$ when the amplitude of oscillations of the forced system is smaller than the amplitude of the external signal. 

When $p=0$, (Fig. \ref{arn}, left panel) the synchronisation region with respect to the frequency $\Omega$ and amplitude $A$ of the external signal has the typical triangular-like shape seen on Arnold tongues.\cite{pik} \ An intermediate fraction of repulsive links ($p=0.43$, Fig. \ref{arn}, right panel) pushes the synchronisation borders beyond the $p=0$ values, allowing for synchronisation of faster signals at weaker forcing.

\begin{figure}
\vspace{0.5cm}
\centerline{\includegraphics[width=9cm]{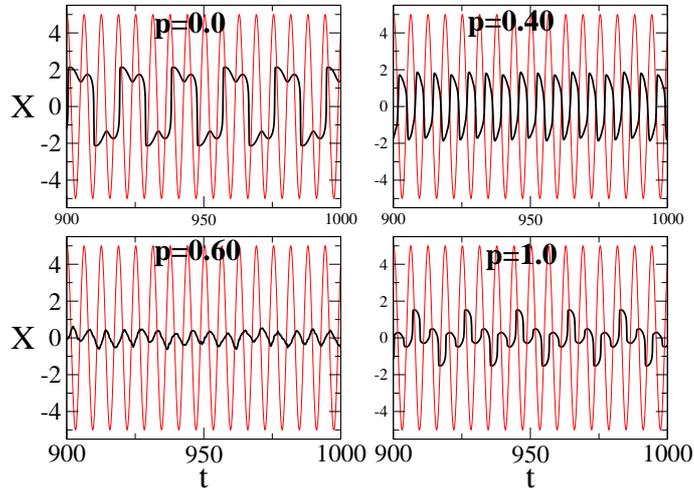}}
\caption{Time evolution of the macroscopic variable $X(t)$ when the system is forced by an external sinusoidal fast signal (lighter color) of amplitude $A=5$ and frequency $\Omega=1.0$, for several probabilities of repulsive links $p$. N, C and $\mu$ as in Fig. \ref{tra}.
}
\label{traf}
\end{figure}

\begin{figure}
\vspace{0.5cm}
\centerline{\includegraphics[width=8cm,angle=0]{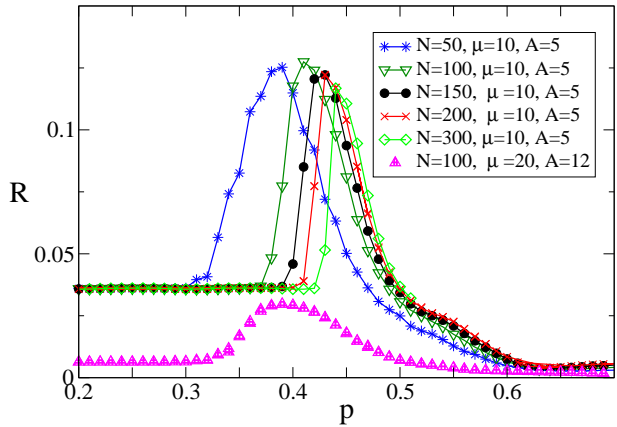}}
\caption{Spectral power amplification, for $C=20$, $\Omega=1$ and several system sizes $N$, averages over 100 runs. We note that smaller systems become synchronised at lower fractions of repulsive links, and are not so dependent on the precise fraction of repulsive links. Additionally, we also observed (figures not shown) a resonance with system size for different probabilities of repulsive links. This kind of dependence has been explained elsewhere.\cite{fi4}
}
\label{spa}
\end{figure}

 Fig. \ref{traf} shows the steady-state trajectory of the macroscopic variable $X(t)$, for different probabilities $p$ of repulsive links, and illustrates the fact a certain proportion of repulsive links is required for the system to adjust its rhythm to that of the external signal (Fig. \ref{traf}, $p=0.40$), whereas  Fig. \ref{spa} confirms that this optimal response only occurs for an intermediate range of the probability of repulsive links. It should be noted that when entrained, the oscillators adjust their frequency while keeping their natural amplitude (Fig. \ref{traf}), therefore, the spectral amplification factor $R$ is smaller than $1$, when $A>2$. As expected, the more the natural frequency deviates from the forcing frequency, the stronger the signal needs to be in order to entrain the system: namely (Fig. \ref{spa}), for a forcing frequency $\Omega=1$, the signal strength needs to be $A=12$ instead of $A=5$, when the natural frequency $\omega=2\pi/T$ is $\approx 0.19$ ($\mu=20$) instead of $\approx 0.39$ ($\mu=10$).

To understand the significance of competitive interactions we recall the results of the last section. The probability region where weak fast signals can be entrained is signalled by the peak in the spectral power amplification $R$, and coincides with the localisation region, as given by the peak in $M$ (Fig. \ref{comp}).

This localisation, or controlled disorder, is crucial for an enhanced response to fast signals, for it allows the amplitude of oscillations to be deformed, varying slightly enough to  place some oscillators closer to the fast motion region, thus allowing a jump to another slow region under an external forcing; but not so much that there is a risk they would trigger a chain reaction, unless there is a permanent pulling into some direction - that is, unless the signal is acting.

\subsection{Weak slow signals benefit from very high $p$}
\label{sec:slow}

In the previous section, we chose to measure the enhancement at the collective level, using the macroscopic variable $X(t)$ in our measure of response $R$, Eq. (\ref{R}); that corresponded to a synchronisation with the external forcing at the individual level: the greater the number of entrained oscillators, the greater the response was. 

\begin{figure}
\vspace{1.0cm}
\centerline{\includegraphics[width=8.5cm,angle=0]{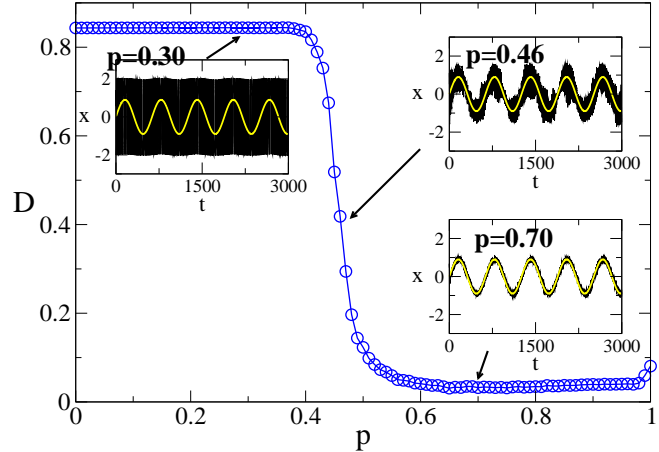}}
\caption{Illustration of representative macroscopic trajectories:  very weak slow signal is best followed the more repulsive connections it has. Other parameters: $A=0.9$, $\Omega=0.01$. N, C and $\mu$ as in Fig. \ref{tra}.
}
\label{slowy}
\end{figure}

\begin{figure}
\vspace{1.0cm}
\centerline{\includegraphics[width=8.5cm]{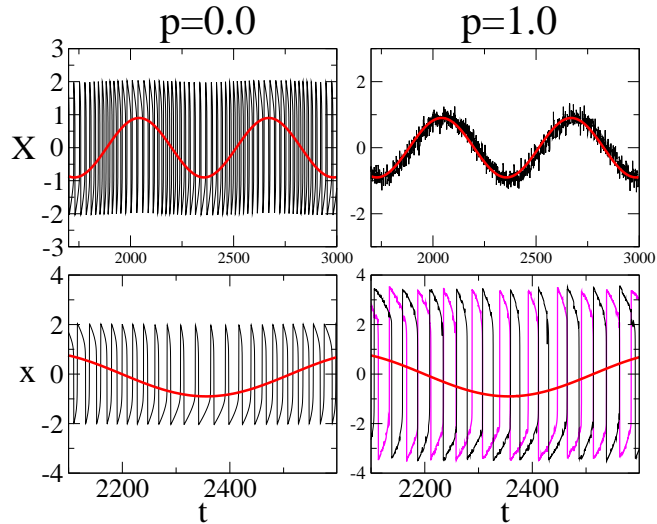}}
\caption{The slower oscillation corresponds to the external signal  while the higher frequency oscillations correspond to either the trajectory of the macroscopic variable $X(t)$ or two typical individual trajectories. Upper panels: When $p=1.0$ we observe the synchronisation of the macroscopic variable  with a signal that is very weak and slow. Lower panels: we zoom, and plot two representative individual trajectories.  $A=0.9$, $\Omega=0.01$. N, C and $\mu$ as in Fig. \ref{tra}.
}
\label{slow}
\end{figure}

We find a different situation when we subject our system to a very weak slow signal with an amplitude that is smaller than $1$, say $A=0.9$. A complete amplitude and frequency synchronisation with this forcing would imply a fast motion, in the interval $[-1, 1]$, without any intercalating period of slow motion, thus basically destroying the defining feature of a relaxational oscillator (Section \ref{model}). We find it impossible for such a weak signal to entrain an individual oscillator. And yet, we observe that for a high enough fraction of repulsive links (insets Fig. \ref{slowy} and Fig. \ref{slow}), there is a nearly coincidence between the trajectory of the global variable $X(t)$ and the forcing $A\sin(\Omega t)$, with an almost imperceptible phase delay. Therefore, the simplest measure of entrainment, that falls to zero if there is a perfect synchronisation, is:

\begin{equation}
D=\frac{\left\langle\left[X(t) - A\sin(\Omega t)\right]^2\right\rangle} {\langle X(t)^2\rangle},
\end{equation}
The results plotted in Fig. \ref{slowy} show how the synchronisation with the external signal as the fraction $p$ of repulsive links mirrors the loss of synchronisation seen in Fig. \ref{dis} for the unforced system. 

Again, the mechanism has its roots on the disorder induced by the presence of repulsive links. As we saw, after the transition region (Fig. \ref{dis}, $p\approx 0.45$), the unforced system tends to adopt a configuration corresponding to a zero value of the macroscopic variable $X$ (Fig. \ref{traf}, $p=0.6$ and $p=1.0$). On the other hand, the forcing induces an asymmetry in the oscillations favouring the time spent on the side of the signal's extremum, as seen in the right panel of Fig. \ref{slow}. Slight as this asymmetry may be, it is enough to cause the superposition of the individual waves on the same side as the signal's extremum, while on the other side, the oscillations cancel each other (see right panel of Fig. \ref{slow} for $p=1.0$). The total synchronisation of the system in case $p$ is zero or small, naturally prevents this phenomenon to happen (see right panel of Fig. \ref{slow} for $p=0.0$).

\section{Further applications: FitzHugh-Nagumo}
\label{ext}

The single uncoupled van der Pol oscillator can  be transformed either into a linear oscillator by taking $\mu=0$ or by replacing the nonlinear damping term $\mu(1 - x_i^2)$, Eq. (\ref{vdp}) by a constant;  or into an excitable system - a simplified FitzHugh-Nagumo - by adding a constant $a$ such that $|a|>1$ to Eq. (\ref{li2}). So, a first direct extension consists in a brief exploration of how those transformations affect our results. 

Not surprisingly, we didn't find an enhanced response for linear oscillators: both mechanisms of enhancement for slow and fast signals rely on the existence of a slow motion region. This situation contrasts with the case studied in.\cite{linosc} \ In that paper, the authors studied a system of linear oscillators with a distribution of natural frequencies. Defining as a measure of diversity the variance of the natural frequencies, they found an optimal response to an external signal for an intermediate level of diversity. Interestingly enough, the enhancement of response also had its origins in an intermediate level of disorder. However, the microscopic mechanism was rather different: some oscillators had a natural frequency that resonated with the signal's frequency, and where able to pull the others due to the positive coupling. In our case, there isn't any single oscillator whose frequency can be entrained by the external signal.

On the other hand, the mechanisms we proposed should be applicable to the FitzHugh-Nagumo model. Adding a constant $a$ to Eq. (\ref{li2}) the system becomes:

\begin{subequations}
\begin{eqnarray}
\dot{ x_i } & = & \displaystyle   \mu \left[x_i - \frac{1}{3}x_i^3 - y_i + \frac D N \sum_{j=1}^N J'_{ij}x_j\right] \label{fhn1}\\
\dot{ y_i } & = & \displaystyle   \frac{1}{\mu} \left[x_i - A \sin (\Omega t)  + a \right]. \label{fhn2}
\end{eqnarray}
\end{subequations}

\begin{figure}
\vspace{0.5cm}
\centerline{\includegraphics[width=8.5cm,angle=0]{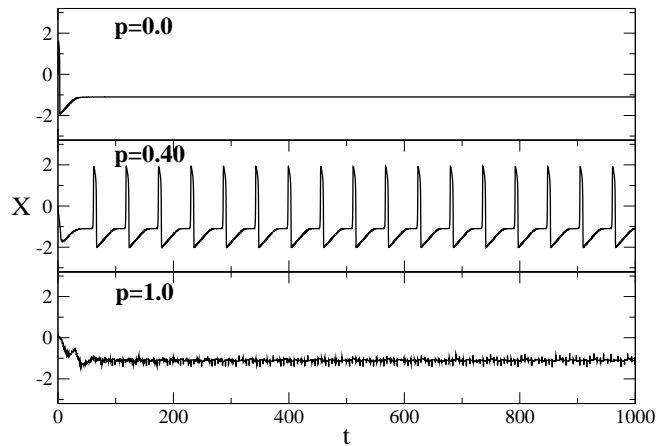}}
\caption{The trajectory of the global variable $X$ for different fractions of repulsive links $p$ in the unforced FitzHugh-Nagumo system showing a similar phenomenon to coherence resonance. Other parameters: $a=1.1$, N, C and $\mu$ as in Fig. \ref{tra}.
}
\label{fhn}
\end{figure}

The interaction via competitive interactions can play the same role of noise or diversity, thus enabling rhythmic excursions away from the fixed point. The result shown in Fig. \ref{fhn} bears  some similarity with the phenomenon by which we observe that the periodicity of oscillations becomes maximally ordered for an intermediate level of noise,\cite{cr, cr1, cr2} diversity,\cite{rot1} or competitive interactions.\cite{Levya, almen} \ In our case, however, and like it was observed in \cite{rot} for the case of active rotators, we don't observe any oscillations at all unless some interactions are repulsive.

\begin{figure}
\vspace{0.8cm}
\centerline{\includegraphics[width=8.5cm,angle=0]{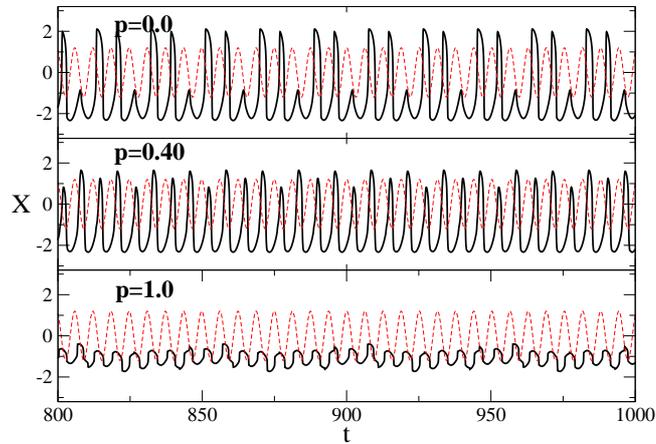}}
\caption{Trajectories of the global variable $X$ for a system of FitzHugh-Nagumo units in the excitable regime forced by a fast signal, for increasing fractions of repulsive links $p$. Other parameters: $a=1.1$, A=12, $\Omega=1$, N, C and $\mu$ as in Fig. \ref{tra}. The pointed line shows the external signal multiplied by $0.1$, for better viewing.
}
\label{fastf}
\end{figure}

When we force the excitable system by a strong enough fast signal (Fig. \ref{fastf}), it starts to oscillate even for $p=0$, and for an intermediate amount of repulsive interactions the main frequency of oscillations coincides with the external signal (Fig. \ref{fastf}, $p=0.40$). 

\begin{figure}
\vspace{0.8cm}
\centerline{\includegraphics[width=8.5cm,angle=0]{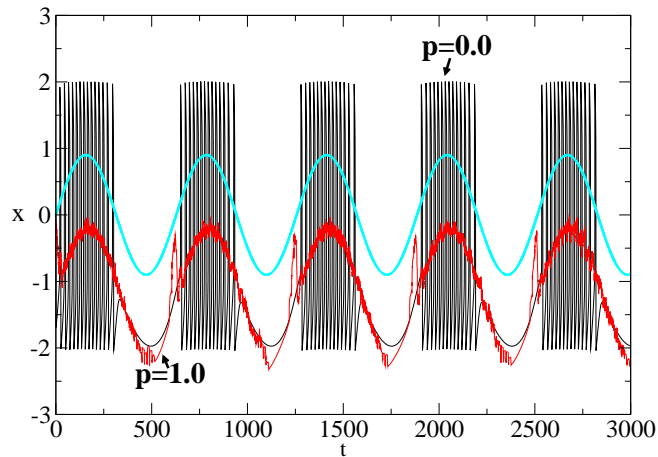}}
\caption{Representative trajectories of the global variable for a system of FitzHugh-Nagumo units in the excitable regime, $a=1.1$, for some probabilities $p$ of repulsive links. Other parameters:  $\Omega=0.01$, A=0.9, and N, C and $\mu$ as in Fig. \ref{tra}. The signal is represented in a lighter color.
}
\label{slowf}
\end{figure}

On the other hand, when we force a system of FitzHugh-Nagumo elements by slow weak signals (Fig. \ref{slowf}) we observe bursts with the periodicity of the signal for $p=0$, while for $p=1$ the global variable roughly oscillates along with the external forcing. Even though the periodicity of the external signal is detected for all fractions of repulsive links, we can imagine situations where we actually want to replicate the behaviour of the external signal, and that is only possible when the fraction of repulsive links is large enough.

Both of these results are expected taking into account the arguments we gave for the van der Pol oscillator case.

\section{Conclusions}
\label{conclusions}

We have shown that the presence of repulsive links in a system of globally coupled van der Pol oscillators can enhance the response to an external signal. This phenomenon is verified regardless of whether the signal is strong and fast, or weak and slow, and it is in every case directly related to a loss of synchronisation and the existence of a slow motion region, but the {\sl microscopic} mechanism of enhancement is different in each case. 

From the point of view of a strong fast signal, the van der Pol oscillator can be approximated by a bistable system, implying a threshold that is regularly overcome with the help of an intermediate proportion of repulsive links, by means of the deformation of the slow motion region.
In the case of very slow signals, the mechanism is associated with the tendentiously anti-phase oscillations that occur when there is a majority of repulsive links.

In both cases, enhancement is directly related to a loss of full synchronisation when the fraction of repulsive links increases. One can imagine that a different coupling scheme might enhance or hinder the results we found, since it is known that the network topology plays a role in synchronisation properties.\cite{arenas}

\section*{Acknowledgements}
We acknowledge financial support by the MICINN (Spain) and FEDER (EU) through project FIS2007-60327 (FISICOS). TVM is supported by FCT (Portugal) through Grant No.SFRH/BD/23709/2005. 

%


\begin{thebibliography}{99}
  
\bibitem{pik} A. Pikovsky, M. Rosenblum and J\"urgen Kurths. {\em Synchronisation: A Universal Concept in Nonlinear Sciences}, Cambridge University Press, 2003.

\bibitem{pace} J. Jalife,  J. Physiol. \textbf{356} (1984), 221.

\bibitem{fet} P. van Leeuwen, D. Geue, M. Thiel, D. Cysarz, S. Lange, C. Romano, N. Wessel, J. Kurths and D. H. Gr\"onemeyerab, Proc Natl Acad Sci U S A.  \textbf{106(33)} (2009), 13661.

\bibitem{pnas}T. Nishikawa and A.E. Motter, Proc. Natl. Acad. Sci. \textbf{107} (2010), 10342.


\bibitem{Levya} I. Leyva, I. Sendi{\~n}a-Nadal, J. A. Almendral, and M. A. Sanjuan, Phys. Rev. E  \textbf{74} (2006), 056112.


\bibitem{tsim} L. S. Tsimring,  N. F. Rulkov,  M. L. Larsen  and M. Gabbay,  Phys. Rev. Lett. \textbf{95} (2005), 014101.


\bibitem{glassy2} L. L. Bonilla, C. P\'erez-Vicente and J.M. Rub\'i, Journal of Statistical Physics \textbf{70} (1993), 921.


\bibitem{daido1} H. Daido, Prog. Theor. Phys. \textbf{77} (1987), 622.

\bibitem{daido2} H. Daido, Phys. Rev. Lett. \textbf{68} (1992), 1073.

\bibitem{daido3} H. Daido, Phys. Rev. E \textbf{61} (2000), 2145.

\bibitem{beam}  N. F. Rulkov, L. Tsimring, M. L. Larsen and M. Gabbay, Phys. Rev. E \textbf{74} (2006), 056205.


\bibitem{preprint} M. Giver, Z. Jabeen and B. Chakraborty,  arXiv:1009.6004.

\bibitem{zan} D. H. Zanette, Europhys. Lett. \textbf{72} (2005), 190.

\bibitem{vdp} B. van der Pol and J. van der Mark, London, Edinburgh, and Dublin Philosoph. Mag., and J. Sci.,  Ser. 7, \textbf{6} (1928), 763.

\bibitem{msf} L. M. Pecora  and  L. T. Carroll, Phys. Rev. Lett. \textbf{80} (1998), 2109. 

\bibitem{pat} P. N. McGraw and M. Menzinger, Phys. Rev. E {\bf 77} (2008), 031102.

\bibitem{fi4} T. Vaz Martins,  V. N. Livina,  A. P. Majtey,  and R. Toral,  Phys. Rev. E \textbf{81}  (2010), 041103.

\bibitem{Perc} M. Perc, M. Gosak and S. Kralj, Soft Matter {\bf 4} (2008), 1861.

\bibitem{and58} P. W. Anderson,  Phys. Rev. {\bf 109} (1958), 1492.

\bibitem{and70} P. W. Anderson, Mater. Res. Bull. {\bf 5} (1970), 549.


\bibitem{G} L. Gammaitoni, P. H\"anggi, P.Jung and F. Marchesoni, Rev. Mod. Phys. {\bf 70}, (1998), 223.

\bibitem{HM} Topical Issue on Stochastic Resonance, Eur. Phys. J. B  {\bf 69}, No. 1, (2009), edited by P. H\"anggi and F. Marchesoni.


\bibitem{TMTG06} C. Tessone, C.R. Mirasso, R. Toral and J.D. Gunton, Phys. Rev. Lett. {\bf 97} (2006), 194101.

\bibitem{TTVL} R. Toral, C. J. Tessone and J. Viana Lopes, Eur. Phys. J. Special Topics {\bf 143} (2007), 59.


\bibitem{linosc} R. Toral, E. Hernandez-Garcia and J. D. Gunton, International Journal of Bifurcation and Chaos {\bf19(10)}  (2009), 3499.

\bibitem{dac} T. Vaz Martins, R. Toral and M.A. Santos, Eur. Phys. J. B  {\bf 67}, N. 3 (2009), 329.


\bibitem{tessone} C. J. Tessone and R. Toral, Physica A \textbf{351} (2005), 106. 


\bibitem{jung} P. Jung and G. Mayer-Kress, Phys. Rev. Lett. \textbf{74} (1995), 2130.

\bibitem{lindner} J. F. Lindner, B. K. Meadows, W. L. Ditto, M. E. Inchiosa and A. R. Bulsara, Phys. Rev. Lett. \textbf{75}  (1995), 3.


\bibitem{kura} Y.~Kuramoto. {\em Chemical oscillations, waves and turbulence}, Springer--Verlag, New York (USA), 1984.

\bibitem{ampli} P. Jung and P. H\"anggi, Europhys. Lett. \textbf{8} (1989), 505.


\bibitem{local}  J. Restrepo, E. Ott, B. Hunt, Phys. Rev. Lett. \textbf{93} (2004), 114101.


\bibitem{cr} A. S. Pikovsky and J. Kurths, Phys. Rev. Lett. \textbf{78} (1997), 775.


\bibitem{cr1} A. Neiman, L. Schimansky-Geier, A. Cornell-Bell  and  F. Moss,  Phys. Rev. Lett. \textbf{83} (1999), 4896.


\bibitem{cr2} C. Kurrer  and K. Schulten,  Phys. Rev. E \textbf{51} (1995), 6213.


\bibitem{rot1} C.J. Tessone, A. Scir\`e, R. Toral and P. Colet, Physical Review E  \textbf{75} (2007), 016203.


\bibitem{rot} C.J. Tessone, D.H. Zanette and R. Toral, The European Physical Journal B \textbf{62} (2008), 319.

\bibitem{almen} J. A. Almendral,  I. Leyva and I. Sendi{\~n}a-Nadal, International Journal of Bifurcation and Chaos \textbf{19(2)} (2009), 711.


\bibitem{arenas} A. Arenas, A. D\'iaz-Guilera, J. Kurths, Y. Moreno and C. Zhou, Physics Reports  \textbf{469} (2008), 93.



\end{thebibliography}
\end{document}